\documentclass[aps,twocolumn,prX,groupedaddress,amsmath,floatfix,showpacs]{revtex4}
\begin{document}

\title{Linear and Non-linear Susceptibilities from Diffusion Quantum Monte Carlo:
Application to Periodic Hydrogen Chains}
\author{P. Umari,$^{1,2}$ and Nicola Marzari$^2$}
\affiliation{$^1$CNR-INFM Democritos, Theory@Elettra Group, Basovizza,
             Trieste, Italy}
\affiliation{$^2$Department of Materials Science and Engineering, MIT, 02139 Cambridge MA}

\date{\today}

\begin{abstract}
We calculate the linear and non-linear
susceptibilities of periodic longitudinal chains of hydrogen dimers with different
bond-length alternations using a diffusion quantum Monte Carlo approach.
These quantities are derived from the changes in electronic polarization as a function of 
applied finite electric field - an approach we recently introduced and made possible by the use of a 
Berry-phase, many-body electric-enthalpy functional.
Calculated susceptibilities and hyper-susceptibilities are found to be 
in excellent agreement with the best estimates available from
quantum chemistry - usually extrapolations to the infinite-chain limit of
calculations for chains of finite length.
It is found that while exchange effects dominate the proper description of 
the susceptibilities, second hyper-susceptibilities are greatly
affected by  electronic correlations. We also 
assess how different approximations to the nodal 
surface of the many-body wavefunction affect the accuracy of 
the calculated  susceptibilities.

\end{abstract}

\pacs{
36.20.-r, 
77.22.-d, 
71.15.-m,  
31.15.A- 
}

\maketitle
\section{Introduction}

The linear and non-linear longitudinal dielectric susceptibilities of periodic 
linear chains of hydrogen dimers have been in recent years the subject of  many density-functional theory
and quantum chemistry studies \cite{cmv95,cm96,nyk98,gsg99,fbl02,ggb02,mwy03,kkp04,bn05,bbl05,psb08,rpc08,kmk08}.
When dielectric properties are considered, these model systems 
present appealing similarities with real polymeric chains.
In particular, in both systems the second hyper-susceptibilities are very large.
Despite the apparent simplicity of such systems, obtaining reliable results
for susceptibilities and hyper-susceptibilities
requires particular care \cite{notefirst}. 
Density-functional theory approaches 
using  the local density (LDA) or generalized-gradient (GGA) approximations
strongly overestimate dielectric response, up to several
orders of magnitude in the case of  the second hyper-susceptibility,
and more sophisticated approaches need to be considered \cite{kk08}.
Correlated quantum chemistry approaches, such as
M\o ller-Plesset (MP) and coupled-cluster (CC),
provide benchmark results, although carefully converged results often require the
most sophisticated levels of approximations (MP4,CCSD(T)) and extended basis sets\cite{cm96}.

Recently, we have introduced a novel scheme for the treatment of finite static homogeneous
electric fields in periodic quantum Monte Carlo (QMC) calculations\cite{uwg05}.  
The use of periodic boundary conditions (PBCs) permits to
avoid the extrapolations to the infinite chain limit, which would be 
even more cumbersome in this case due to the presence of QMC statistical errors.
When applied to the hydrogen chain model, this approach provides
linear susceptibilities in excellent agreement with the results
from the most accurate quantum chemistry calculations.
Since it has been shown that, in contrast to the linear response case, electronic
correlations account for a large extent to second hyper-susceptibilities\cite{cm96},
it is of particular interest to assess the accuracy and  reliability
of QMC for the calculation of these higher-order responses.
Indeed in QMC, in the limit of a correct nodal surface approximation, 
electronic correlations are treated exactly and basis sets approximations are 
easily controlled.  Moreover, due to the favorable scaling of QMC approaches with respect
to the number of electrons (cubic, or even linear \cite{whg01}), this
approach could represent a very viable strategy for accurate, predictive calculations of 
non-linear dielectric properties in realistic systems.

Here, we calculate with diffusion quantum Monte Carlo (DMC) 
the linear polarizability and second hyper-polarizability per H$_2$ unit
(i.e. the susceptibilities) for the paradigmatic case of periodic linear chains of 
hydrogen dimers, where the dimers have a fixed length of 2.0 a.\ u.\, and we consider
three different bond-length alternations of 2.5 a.\ u.\, 3.0 a.\ u.\ and 4.0 a.\ u.\ respectively.

The paper is organized as follows: Section \ref{section2} briefly describes our methodology to
treat finite electric fields in periodic QMC calculations. In Section \ref{section3} we summarize
the technical details of the calculations and investigate the degree of convergence 
with respect to the different approximations used.
The final results are reported and discussed in Section \ref{section4}, while
in Section \ref{section5} we investigate the errors induced by different
choices for the many-body nodal surface.
Conclusions, and perspectives for future work are drawn in Section \ref{section6}.

\section{Method}
\label{section2}

We briefly summarize here the approach we first introduced in Ref.\ [\onlinecite{uwg05}].
We consider a system of $N$ interacting electrons in a periodic
cell of size $L$ (to simplify the notation we describe here the
one-dimensional case, but the extension to higher dimensions is
straightforward).
The normalized many-body wavefunction $\Psi$ obeys  PBCs
\begin{equation}
\Psi(x_1,..,x_i+L,..,x_N)=\Psi(x_1,..,x_i,..,x_N),
\end{equation}
for every $i$.
On the other hand, the position operator
\begin{equation}
\hat{X}=x_1+x_2+...+x_N,
\end{equation}
does not satisfy PBCs and is unbounded from below; thus it cannot be used (and would
be ill-defined) to calculate expectation values, say, of the potential generated by a constant electric field.
On the other hand, the modern theory of the polarization\cite{kv93,wm01,r92} (MTP) and its
many-body generalization \cite{r98} provide
definitions for the relevant observables which remains compatible with PBCs.
Indeed, the polarization $P$ of a many-body system can be obtained from
the single-point Berry phase \cite{r98} formulation
\begin{eqnarray}
\label{pola}
P[\Psi]&=&-\frac{e}{\Omega}\frac{L}{2\pi} {\mathrm{ Im}} \ln z\\
\label{zeta}
z[\Psi]&=&\langle\Psi|e^{iG\hat{X}}|\Psi\rangle,
\end{eqnarray}
where $\Omega$ is the size of the cell (in 1d is equal to $L$),
$e$ is the elementary charge, $G=2\pi/L$, and $z$ will be termed here as the
complex polarization.
The definition of Eq.\ (\ref{pola}) coincides, in the thermodynamic limit,
with the exact many-body observable \cite{r98}, but it remains well-defined for
any finite $L$. Although in Eq.\ (\ref{pola}) the polarization
is defined modulo $-eL/\Omega$, the quantities which are experimentally observed
correspond to differences of polarization and are well defined within the MTP.

Having a valid definition for the many-body polarization, we can operate
a Legendre transform of our total energy functional between the conjugate variables of 
electric field ${\mathcal E}$ and polarization $P$, and obtain the
ground-state wavefunction of a periodic, extended system
in the presence of an electric field $\mathcal{E}$
from the minimum of the generalized electric enthalpy
\cite{nv94,ng01,up02,siv02,sbp04}
\begin{equation}
\label{entalpia}
F[\Psi]=E^0[\Psi]-{\mathcal E}\Omega P[\Psi],
\end{equation}
where $E^0[\Psi]$ is the energy functional for the unperturbed Hamiltonian $H^{0}$.
The direct minimization of an electric enthalpy functional  has been
firstly introduced for tight-binding Hamiltonians in Ref. [\onlinecite{nv94}]
and in the context of density functional theory in Refs.\ [\onlinecite{up02,siv02}]. 
On the other hand, DMC and variational Monte Carlo (VMC) (with variance minimization approaches)
require the use of Hamiltonian  operators which are both local and Hermitian.
This requirement can be satisfied by noting that
a local Hermitian operator 
can be identified 
from the minimum condition for the electric enthalpy
\begin{equation}
\frac{\delta F}{\delta\langle\Psi|}=\lambda|\Psi\rangle ,
\end{equation}
with $\lambda$  the appropriate Lagrange multiplier. 
This strategy has already been used for the case
of single-particle Slater determinants \cite{siv04}
and model Hubbard Hamiltonians \cite{sbp04}.
This local Hermitian operator shares the same ground-state wavefunction of
the electric enthalpy functional, and
the $\Psi$ that minimizes Eq.\ (\ref{entalpia})
is also the ground-state for the  many-body Hamiltonian
\begin{equation}
\label{hamilt}
H(z)=H^{0}+{\mathcal E}\frac{eL}{2\pi}
{\mathrm{Im}}\frac{e^{iG\hat{X}}}{z}
\end{equation}
It should be noted that Eq. (\ref{hamilt}) defines a {\it self-consistent many-body Hamiltonian},
since it's an operator that depends on the complex polarization $z$, and thus on the
expectation value of $\Psi$ through Eq.\ (\ref{zeta}). Self-consistency is indeed the price to be paid to
recast the variational principles of the electric-enthalpy functional into an operator equation.

Due to the self-consistent nature of the operator $H(z)$ defined in Eq.\ (\ref{hamilt}),
the ground state in the presence of an electric field must be found through
an iterative procedure. We start from a first value $z_1$ for $z$, e.g. as found in
the single-particle calculations or in the many-body trial wave function $\Phi_{\mathrm T}$;
the local Hamiltonian $H(z_1)$ is then constructed. DMC evolution using this operator leads
to a new expectation value for $z$, called
$z_2$, which in turn determines a second Hamiltonian $H(z_2)$.
In the absence of stochastic noise, this process could be iterated to convergence:
\begin{equation}
\label{series_z}
z_1\rightarrow z_2\rightarrow z_3\rightarrow...\rightarrow z_n,
\end{equation}
to find the fixed point of the complex-plane map
\begin{equation}
\label{effe}
f(z_i)=z_{i+1}.
\end{equation}
Since the Monte Carlo procedure introduces a statistical error
in every estimate of $z_i$,
the map $f$ becomes a stochastic function in the complex plane.
If $f$ can be approximated close to its fixed point as
a linear function in the complex plane, it has been shown\cite{uwg05} that
its average over a sequence of $\{z_i\}$ provides the best estimate
for the fixed point itself.
The validity of this approximation must be checked numerically.
Once the fixed point $\overline z$ is obtained, the corresponding
polarization is then obtained via Eq.\ (\ref{pola}).

\section{Computational details}
\label{section3}

We investigate  in this study dielectric properties of 
periodic linear chains of hydrogen dimers along the longitudinal direction.
The chosen dimer bond length is 2.0 a.\ u.\ and bond-length alternations 
of 2.5, 3.0 and 4.0 a.\ u.\ are considered.
Because of symmetry, these systems have vanishing  first hyper-susceptibility.
We treat these systems through periodic boundary conditions.
We first perform  HF calculations using an orthonormal unit cell 
with an edge size of 20 a.\ u.\ in the directions perpendicular to
the chain and  containing one single H$_2$ unit.
We sample the one-dimensional Brillouin zone along the  chain
direction with a set of $N_k$ equally spaced k-points.
Then, for QMC calculations the corresponding supercell containing
$N_k$ hydrogen dimers, sampled at the $\Gamma$-point, is
considered.
The HF calculations are performed using the {\sc pwscf}
package from the {\sc Quantum-ESPRESSO} distribution \cite{pwscf}.
The wave-functions are expanded in a plane-wavebasis set
with a cutoff of 50 Ry.
Finite electric fields are implemented through the algorithm 
described in Ref.\ [\onlinecite{siv04}].
To describe the Coulomb potential of the H nucleus we use
a norm-conserving pseudopotential. 
For this same pseudopotential we already showed in
Ref.\ [\onlinecite{uwg05}] that
the calculated DMC polarizability of the isolated H atom is
in perfect agreement with the theoretical value.

The convergence of the Berry-phase polarization to the thermodynamic limit
is of the order $L^{-2}$ with respect to the supercell dimension $L$; this
can be understood on purely geometrical grounds (see Ref. [\onlinecite{up03}]). Thus,
we determine here the k-point sampling $N_k$ for the primitive unit cell, and equivalently the 
length of a supercell sampled at the $\Gamma$ point, for which the sampling/geometrical
errors in the calculated linear susceptibility and second hyper-susceptibility become smaller 
or negligible with respect to the magnitude of the statistical errors due to the DMC procedure.
We investigate this convergence within the HF approach,
focusing on chains with a bond-length alternations of 3 a.\ u.\ ,
for which also extrapolations to the infinite chain limit
for both the polarizability and hyper-polarizability
have been reported for several basis sets in Ref.\ [\onlinecite{cm96}].
We consider two values for $N_k$, 10 and 20, and use two applied electric
fields of magnitude 0.003 and 0.02 a.\ u.\ respectively.
We then fit the dipole moments $\mu$  per H$_2$ unit with 
\begin{equation}
\label{fit}
\mu(\mathcal{E})=\alpha\mathcal{E}+\frac{1}{6}\gamma\mathcal{E}^3,
\end{equation}
where $\mathcal{E}$ is the electric field and $\alpha$ and $\gamma$ are the linear susceptibility
and the second hyper-susceptibility per H$_2$ unit.
We report in Tab.\ \ref{tab1} the values for $\alpha$ and $\gamma$ we calculated using PBCs, together
with the extrapolations for the most complete basis sets presented in Refs.\ [\onlinecite{cmv95,cm96}],
showing very good agreement with these published calculations.
As we want to focus here on the non-linear susceptibility we will consider in the following
supercells of 10 dimers each, i.e. corresponding to $N_k=10$ sampling.

The calculated HF wavefunctions are then interpolated with splines 
and imported into the {\sc CASINO} VMC and DMC codes \cite{casino}.
In VMC the $N$-body wavefunction $\Psi_\mathrm{VMC}$ is defined as
\begin{equation}
\label{vmc_wfc}
\Psi_\mathrm{VMC}(\mathbf{r}_1,..,\mathbf{r}_N)=exp(J(\mathbf{r}_1,..,\mathbf{r}_N))\frac{1}{\sqrt{N!}}\det
\left(\psi_1..\psi_N\right),
\end{equation}
where $\{\mathbf{r}\}$ are the electron positions, $\{\psi\}$ are the HF single-electron
wavefunctions and $J$ is the Jastrow factor. In this work we use the  Jastrow factor formulation
introduced in Ref.\ [\onlinecite{fmn01}], which depends only on the distances 
$|\mathbf{r}_i-\mathbf{r}_j|$ and $|\mathbf{r}_i-\mathbf{R}_I|$, where $\{\mathbf{R}\}$ are 
the ionic positions.
We use an expansion to the sixth order for both spin components, and two different
sets of parameters for inequivalent H ions. Indeed, when a longitudinal field
is applied, the two H atoms in the unit cell are no longer equivalent.
To optimize the Jastrow parameters we apply a variance minimization scheme\cite{uww88}.
For VMC simulations we adjust the time-step in order to assure 
an acceptance ratio of $\sim$50\%.
It should be noted that we use a spherical symmetrical Jastrow factor that depends
only on the  relative distance between electrons and between electrons and nuclei.
Such form does not allow, in the VMC procedure, for significant changes of the expectation value 
of $z$ (defined in Eq.\ (\ref{pola})) with respect to its original HF value.
Moreover, the variance minimization scheme used here is less sensitive to 
long range properties than energy minimization schemes \cite{lzr00,uf05}

In the following, we consider
a hydrogen chain with a bond-length alternation of 2.5 a.\ u.\ ,
as done in Ref.\ [\onlinecite{uwg05}]. 
We report in Tab.\ \ref{tab2}
the estimates for the total electric enthalpy and $z$ complex polarization
obtained in the presence of 
an electric field of 0.003 a.\ u.\
and for several consecutive steps of the Jastrow factor optimization. 
As expected,  we note a significant lowering of the total electric enthalpy due 
to the $E^{0}$ term in Eq.\ (\ref{entalpia}), while the polarization
remains almost unchanged.

Once the Jastrow factor has been optimized, the resulting $N$-body wavefunction
$\Psi_\mathrm{VMC}$ is used as trial wavefunction for the DMC simulations,
which are based on importance sampling. 
In all the DMC calculations we use a time step of 0.02 a.\ u.\ assuring
an acceptance ratio greater than 99.7\%.
Indeed, in the DMC simulations
the walkers are distributed as $\Psi_\mathrm{VMC}\Psi$, where $\Psi$ is the correct
$N$-body ground-state wavefunction compatible with the nodal surface defined
by $\Psi_\mathrm{VMC}$. The expectation value $\overline{O}$ of an operator $\hat{O}$  which
does not commute with the Hamiltonian and depends  only on the electron
positions $\{\mathbf{r}\}$ is obtained 
through a forward-walking procedure \cite{lkc74,hlr94}
\begin{equation}
\label{forward}
\overline{O}=\frac{\sum_\tau\sum_{j=1,N_\tau} O(\{\mathbf{r}_{j,\tau-\Delta t}\}}
{\sum_\tau N_\tau},
\end{equation}
where $N_\tau$ is the number of walkers at time-step $\tau$
and $\mathbf{r}_{j,\tau-\Delta t}$ corresponds
to the ancestor configuration of the walker
$j$ at an interval $\Delta t$ back in imaginary time.
To determine the projection time $\Delta t$ we start from a $\Psi_\mathrm{VMC}$
calculated for an electric field of 0.003 a.\ u.\ and switch off the electric
field $\mathcal{E}$ in the following DMC simulation. Then we calculate the expectation values of
$z$ for increasing values for $\Delta t$ and through Eq.\ (\ref{pola}) the corresponding dipole moments.
These are expected to decay exponentially toward a small value.
Indeed constraining   $\Psi$ to the nodal surface of  $\Psi_\mathrm{VMC}$, 
which is calculated for a non-zero  electric field, prevents the complete relaxation 
of the dipole moment when the field is switched off in the DMC simulation.
In Tab.\ \ref{tab3} we display the decay of the dipole moment for $\Delta t$ ranging from
0 to 1000 time steps. In this study we then use 
a $\Delta t$ corresponding to 1000 time steps which assures the convergence 
of the estimated values of $z$ with respect to the statistical errors of our DMC simulations.
We do not report in the calculated DMC errors the residual bias  
due to the use of the finite projection time $\Delta$.

Last, to verify the linearity of the map $f$ in Eq.\ (\ref{effe}) in the vicinity
of its fixed point we perform two different 
sequences of DMC simulations, as described in Eq. (\ref{series_z}), starting
with the $z$ value given by HF in an applied electric field
of 0.003 a.\ u.\ . While both series have a population of 2560 walkers,
the first one is composed of 20 DMC simulations of 40000 time steps
each, while  the second  one is composed of 6  DMC simulations of 120000  time steps
each. Therefore, larger error bars for each estimate $z$ 
are found in the first series. We expect that if $f$ is linear the calculated average
values $\overline z$ should be the same for the first and second series,
within the statistical error.
Indeed, we report in Tab. \ref{tab4} the average  $\overline z$ together with
the corresponding dipole moments found for the two series. Both quantities 
are found to be equal within the statistical errors. 
Thus, in the following DMC simulations we use 
runs of 40000 time steps each and with a population of 2560 walkers.

\section{Results and discussion}
\label{section4}

Using the computational scheme and optimal parameters described in the previous
sections we consider linear hydrogen chains with bond-length alternations  
of 2.5, 3.0 and 4.0 a.\ u.\ , intermolecular distances of 2.0 a.\ u.\, described with 
10-dimer
supercells in PBCs, to calculate the DMC best estimates for the
fixed-point $\overline z$ at two different values of the applied electric field.
We report in Tab.\ \ref{tab5} these
estimates $\overline z$ for the complex polarization, together with the corresponding dipole moments
per H$_2$ unit.  Since these systems
are centrosymmetric, we consider the dipole moment to be zero in the absence an electric field. 
For larger  bond-length  alternations  it is  possible to access 
larger field intensities without runaway solutions in the HF calculations \cite{up02}.

Then, we fit the calculated moments with the expansion of Eq.\ (\ref{fit}),
and obtain the DMC estimates for the linear susceptibility $\alpha$ and
second hyper-susceptibility $\gamma$. The linear susceptibilities 
are reported in Tab.\ \ref{tab6} together
with the quantum chemistry extrapolations of Ref. [\onlinecite{cmv95}].
We determine $\alpha$ with statistical errors varying from 1.5 \% to 0.4 \%
for the 3.0 a.\ u.\ and 4.0 a.\ u.\ bond-length alternations, respectively. 
The higher precision in the latter
case is due to the application of  higher electric fields.
The reported M\o ller-Plesset susceptibilities calculated for the levels of
approximation MP3 and MP4 and for the  
(6)-31G(*)* and (6)-311G(*)* basis sets show a good degree of convergence 
with the level of approximation (from MP3 to MP4)
and a slightly worse  degree of convergence when using a more complete basis set
( from (6)-31G(*)* to (6)-311G(*)*).
Similar results to the MP4 case are  obtained using the coupled cluster CCSD(T) 
method, as illustrated in Ref.\ [\onlinecite{cmv95}].
Our DMC estimates for $\alpha$ are in excellent agreement with the MP4 values
calculated for the basis set  (6)-311G(*)*. 
In Ref.\ [\onlinecite{uwg05}], the same good agreement 
was found for the chain with a bond-length alternation of 2.5 a.\ u.\ and
using a  trial wavefunction obtained from a density-functional calculation.

We show in Tab.\ \ref{tab7} the calculated DMC estimates for the second 
hyper-susceptibility $\gamma$, together with the 
quantum chemistry extrapolations of Ref. [\onlinecite{cm96}].
We  determine $\gamma$ with statistical errors varying from 6.8 \%
to 3.6 \% for the 3.0 a.\ u.\ and 4.0 a.\ u.\ bond-length alternations, respectively.
Values from MP quantum chemistry extrapolations are available only for the
3.0 a.\ u.\ bond-length alternation case. These show slow convergence
in the MP series (MP3 to MP4) and are still not converged with
the basis set (from (6)-31G(*)* to (6)-311G(*)*).
However, in Ref. [\onlinecite{cm96}] it has been shown that
$\gamma$ calculated through MP3 and MP4 increases while the basis set
is becoming more complete.
Therefore, the best MP4 value should be taken as a lower limit.
In fact, the $\gamma$ coefficient calculated through DMC is found to be higher than
the MP4  counterpart by a factor of $\sim$20\%.

By addressing the difference between the susceptibilities  calculated with HF and those
calculated with DMC, we can investigate the relevance of electronic
correlations. 
We report in Tab.\ \ref{tab8} these differences
with respect to the corresponding DMC susceptibilities and hyper-susceptibilities.
Electronic correlations lower the value of linear susceptibilities $\alpha$
and increase that of second hyper-susceptibilities $\gamma$.
As already shown  in Ref.\ [\onlinecite{cm96}], the effects of correlations are
large for the second hyper-susceptibility $\gamma$. Indeed, they account for almost
half of the calculated DMC value for the chain with a bond-length
alternation of 2.5 a.\ u.\ . As the bond-length alternation increases
the correlation  contribution becomes smaller but still remains considerable.
Therefore,  when calculating non-linear  susceptibilities in such systems,
an adequate treatment of electronic correlations is mandatory and 
QMC approaches are particularly appealing since  in
the limit of the exact nodal surface they treat correlations exactly.

It is worth investigating the localization of the $N$-body wavefunction $\Psi$
along the longitudinal chain direction. The MTP 
provides a definition of the localization spread $\sigma^2$
which remain valid  also within PBCs\cite{rs99}. 
This depends solely on the complex number $\overline z$ which
we estimated in DMC, and the following holds

\begin{equation}
\label{localization}
\sigma^2=-\frac{L^2}{N4\pi^2}\ln|z|^2.
\end{equation}
We report in Tab.\ \ref{tab9} the DMC spreads $\sigma^2$
for the bond-length alternations and electric fields 
addressed. We note that as the bond-length alternation
increases the system becomes more localized. Indeed in the limit of
large bond-length alternations, the  system becomes composed
of isolated  hydrogen dimers.
The opposite behavior is observed when the electric field is increased.
Indeed an applied electric field closes
the electronic gap, and this determines an increase of the electronic
spread as shown in Ref.\ [\onlinecite{swm00}].

Finally, we address the converge of the calculated DMC linear susceptibilities  and second-hyper susceptibilities
with respect to the size of the supercell used in the simulation. We consider a bond length alternation
of 4 a.\ u.\ and we perform an  additional calculation 
for  a supercell consisting of 20 H$_2$ units. 
We use the same calculation parameters of the previous DMC simulations.
We see from the figures reported in  Tab.\ \ref{tab10}  that, in the limit of the statistical error, many-body effects
play  a  minor role. However, a full assessment of the convergence of the second-hyper susceptibility
with respect to the supercell size would require significant smaller error-bars.

\section{Nodal surface approximation}
\label{section5}

In this section we address the dependence of the calculated DMC
dipole moments with respect to the nodal surface through the
choice of the trial wavefunction.
To estimate the magnitude of the error due to an approximate nodal surface, we
consider the hydrogen chain with a bond-length alternation
of 2.5 a.\ u.\ and proceed in this way: first we calculate 
HF trial wave-functions applying electric field intensities of 0.0, 0.003 and 0.01 a.\ u.\;
then, the following VMC and DMC simulations are performed for an electric field of 0.003 a.\ u.\ and
the fixed points $\overline z$ and dipole moments are compared.
The discrepancies between the DMC values calculated starting from  HF wavefunctions obtained consistently 
for an electric field intensity  of  0.003 a.\ u.\ and those from wavefunctions obtained at fields
 of  0. and  0.01 a.\ u.\ are due  to the different nodal surfaces. 
These values are displayed in Tab.\ \ref{tab11} together with the corresponding dipole moments.
We note that for our system the imaginary part of $\overline z$ is most affected 
by the choice of the starting trial wavefunction. As expected, the dipole moment is found to be
 smaller
when starting from  an  HF wavefunction presenting  a smaller dipole moment and becomes
larger when the  HF wavefunction  has a larger one.
However, this discrepancy is not very large: when starting from a 0.\ a.\ u.\ 
dipole moment a difference of  only $\sim$ 7\% is observed.  This discrepancy increases
up to $\sim$ 20\% (i.e. three times larger, but still much smaller than the difference with the initial HF value) 
when starting from the HF wavefunction
with a dipole moment of 0.6055 a.\ u.\ .
We can conclude that the error in the DMC dipole moments is negligible
when using trial wavefunctions calculated for the same values of the
 electric field.
The same conclusion was drawn in Ref.\ [\onlinecite{uwg05}], where it was 
shown that when starting from LDA trial wavefunctions,
which
strongly overestimate the linear susceptibilities,
the DMC linear susceptibility 
was still in excellent agreement with the best quantum chemistry results.
This could be due to the resemblance with genuine one dimensional systems for which the nodal surface 
is fixed by symmetry.

\section{Conclusions and perspectives}
\label{section6}

We have shown how it is possible to calculate linear and non-linear 
susceptibilities of periodic systems using DMC, obtaining for the cases
studied statistical errors lower than 1.5\% for the linear
susceptibilities and  7\% for the non-linear susceptibilities.
The calculated values are in excellent agreement (when available) 
with the results obtained from the most accurate 
quantum chemistry approaches, but do not suffer from basis set
errors or extrapolations to the infinite limit.
These results and approach can be expected to be used as reference for
testing  novel first-principles approaches for the
evaluation of dielectrics properties.
Although DMC calculations of susceptibilities are computational
demanding, they  can benefit, 
in terms of computational speed, from the use 
of order-$N$ methods based on localized 
orthogonal \cite{mv97,whg01} or nonorthogonal basis sets \cite{ag04,rw05}.
Therefore, it  would be of great interest to determine 
the accuracy of the calculated dielectric properties
when such approximations are used.
DMC estimates of linear and non-linear
dielectric properties could be particularly
important for systems for which less expensive approaches, such as 
density-functional theory, fail.
Polymers and conjugated organic molecules are 
a prototypical class of these systems, of great theoretical and
practical interest.

\section*{Acknowledgments}

This research has been supported by the DARPA
the DOE Scidac Institute for Quantum Simulations of Materials
and Nanostructures.
We are especially grateful to Richard Needs and Mike Towler
of the University of Cambridge (UK)
for making their  QMC code {\sc Casino} \cite{casino} available to us.

\bibliographystyle{PhysRev}

\begin{thebibliography}{999}
\bibitem{notefirst}
We consider here symmetric systems for which 
the first hyper-susceptibility vanishes.
\end{thebibliography}

\clearpage

\begin{table}
\begin{tabular}{lcccc}
\hline
& $N_k=10$ & $N_k=20$ & (6)-31G(*)* & (6)-311G(*)* \\
\hline
$\alpha$ & 28.5 & 28.5 & 28.5 & 28.6 \\
$\gamma$ & 56.0 & 57.1 & 55.1 & 56.7 \\
\hline
\end{tabular}
\caption{\label{tab1}
Hartree-Fock linear susceptibility $\alpha$ in a.\ u.\ per H$_2$ unit and second hyper-susceptibility
$\gamma$ in 10$^3$ a.\ u.\ per H$_2$ unit for the periodic linear chain
of  H$_2$ dimers ($d$=2.0 a.\ u.\ ) with a bond-length alternation of 3 a.\ u.\ ,
calculated using PBCs and meshes of 
 $N_k=10$ and  $N_k=20$ 
equally spaced k-points.
The quantum chemistry extrapolations to the infinite limit, obtained 
with basis sets (6)-31G(*)* and  (6)-311G(*)*,
are taken from Refs. [\protect\onlinecite{cmv95,cm96}].
}
\end{table}

\begin{table}
\begin{tabular}{cccc}
\hline
\#VMC run &$F$&$\Re\{z\}$&$\Im\{z\}$\\
\hline
0&-10.6855 $\pm$ 0.0011 &0.6118 $\pm$ 0.0004 &0.1443 $\pm$ 0.0004\\
1&-11.1162 $\pm$ 0.0015 &0.6128 $\pm$ 0.0013 &0.1453 $\pm$ 0.0013\\
2&-11.1608 $\pm$ 0.0014 &0.6128 $\pm$ 0.0014 &0.1472 $\pm$ 0.0014\\
3&-11.1582 $\pm$ 0.0014 &0.6141 $\pm$ 0.0014 &0.1466 $\pm$ 0.0014\\
\hline
\end{tabular}
\caption{\label{tab2}
Electric enthalpy  $F$ and complex polarization $z$ obtained in 4 consecutive VMC runs
for a periodic hydrogen linear chain of H$_2$ dimers ($d$=2.0 a.\ u.\ ) with a bond-length alternation 
of 2.5 a.\ u.\ and an applied electric field of 0.003 a.\ u.\ . The supercell contains 10 dimers.
In the 0-th run no Jastrow term is used and the simulation has
a population of 3.2*10$^6$ walkers. In runs 1-3 
the Jastrow factor is optimized starting from the parameters 
used in Ref.\ [\protect\onlinecite{uwg05}],
and the simulations have a population of 
3.2*10$^5$ walkers.}
\end{table}

\begin{table}
\begin{tabular}{lc}
\hline
$\Delta t$ (time steps) & $\mu$ (a.\ u.\ )\\
\hline
0&30.8   $\pm$ 1.3\\
100&22.8 $\pm$  1.4\\
200&17.6 $\pm$  1.5\\
300&13.9 $\pm$ 1.6\\
400&11.2 $\pm$ 1.7\\
500&9.2 $\pm$  1.8\\
600&7.8 $\pm$ 2.0\\
700&6.8 $\pm$  2.1\\
800&6.1 $\pm$ 2.3\\
900&5.6 $\pm$ 2.4\\
1000&5.4 $\pm$ 2.6\\
\hline
\end{tabular}
\caption{\label{tab3}
DMC dipole moment $\mu$ per H$_2$ unit
for a linear periodic chain (10-dimer supercell) with a bond-length alternation
of 2.5 a.\ u.\ , as a function of the forward-walking projection time $\Delta t$.
The VMC trial wavefunction used has been determined with an applied
electric field of 0.003 a.\ u.\ .
No electric field is applied during the DMC simulations, each involving
1000 walkers and 36000 time steps.}
\end{table}

\begin{table}
\begin{tabular}{llccc}
\hline
\# time-steps & \# runs & $\Re \{\overline z\}$ & $\Im \{\overline z\}$ & $\mu$ (a.\ u.\ )\\
\hline
40000  & 20& 0.6122 $\pm$ 0.0007  & 0.1345 $\pm$ 0.0016 & 0.1549 $\pm$ 0.0020\\
120000 & 6 & 0.6130 $\pm$ 0.0008 & 0.1334 $\pm$ 0.0015 & 0.1545 $\pm$ 0.0018\\
\hline
\end{tabular}
\caption{\label{tab4}
Real and imaginary part of the estimated fixed point $\overline z$ for the
complex polarization and
corresponding dipole moment $\mu$ per H$_2$ unit, calculated
for a linear periodic chain (10-dimer supercell) with a bond-length alternation of 2.5 a.\ u.\
and an applied electric field of 0.003 a.\ u.\ , for two iterative
series with (\# time-steps) time-steps per single run, and (\# runs) 
total number of DMC runs.}
\end{table}

\begin{table}
\begin{tabular}{lcccc}
\hline
$L$(a.\ u.\ ) & $\mathcal{E}$ (a.\ u.\ )& $\Re\{\overline z\}$ & $\Im\{\overline z\}$ & $\mu$ (a.\ u.\ ) \\
\hline
2.5 & 0.003 & 0.6127 $\pm$ 0.0008 & 0.1341 $\pm$ 0.0013 & 0.1544 $\pm$ 0.0018\\
2.5 & 0.01  & 0.3736 $\pm$ 0.0020 & 0.4320 $\pm$ 0.0034 & 0.6143 $\pm$ 0.0050\\
\hline
3.0 & 0.003 & 0.7544 $\pm$ 0.0005 & 0.0775 $\pm$ 0.0013  & 0.0815 $\pm$ 0.0015\\
3.0 & 0.02  & 0.4724 $\pm$ 0.0040 & 0.5162 $\pm$ 0.0021  & 0.6602 $\pm$ 0.0053\\
\hline
4.0 & 0.01  & 0.8460 $\pm$ 0.0002 & 0.1459 $\pm$ 0.0008   & 0.1631 $\pm$ 0.0010\\
4.0 & 0.03  & 0.6996 $\pm$ 0.0008 & 0.4600 $\pm$ 0.0013  & 0.5554 $\pm$ 0.0016\\
\hline
\end{tabular}
\caption{\label{tab5}
Real and imaginary part of the estimated fixed point $\overline z$ for the
complex polarization, and 
corresponding dipole moment $\mu$ per H$_2$ unit,
for periodic hydrogen chains with bond-length alternation $L$
and applied electric field $\mathcal{E}$.}
\end{table}

\begin{table}
\begin{tabular}{lccccc}
\hline
$L$(a.\ u.\ ) & DMC & MP3(a) & MP3(b) & MP4(a) & MP4(b)\\
\hline
2.5    & 50.57 $\pm$ 0.50  &  51.35 & 54.33 & 50.02 & 53.56\\
3.0    & 27.03 $\pm$ 0.55  &  25.66 & 27.01 & 24.94 & 26.51\\
4.0    & 16.04 $\pm$ 0.10 &  15.4  & 16.13 & 15.00 & 15.83\\
\hline
\end{tabular}
\caption{\label{tab6}
DMC linear susceptibility $\alpha$ in a.\ u.\ per H$_2$ unit for
periodic linear hydrogen chains with bond-length alternation $L$,
compared with  quantum chemistry results from Ref.\ [\protect\onlinecite{cmv95}]
 reported for the basis set (6)-31G(*)* (a) and (6)-311G(*)*  (b).}
\end{table}

\begin{table}
\begin{tabular}{lccccc}
\hline
$L$(a.\ u.\ ) & DMC & MP3(a) & MP3(b) & MP4(a) & MP4(b)\\
\hline
2.5    &  652. $\pm$ 61. &   &  &  & \\
3.0    &  89.8 $\pm$ 12. & 65.73 $\pm$ 0.06  &  73.00 $\pm$  0.05 & 65.77 $\pm$ 0.11 & 74.68 $\pm$  0.05\\
4.0    &  16.5 $\pm$ 1.1 &   &  &  & \\
\hline
\end{tabular}
\caption{\label{tab7}
DMC second hyper-susceptibility $\gamma$ in units of
10$^3$ a.\ u.\ per H$_2$ unit for periodic linear hydrogen chains with 
bond-length alternation $L$; for the chain with $L$=3.0 a.\ u.\ ,
quantum chemistry MP results are taken from Ref.\ [\protect\onlinecite{cm96}],
and are reported for the basis sets (6)-31G(*)* (a) and (6)-311G(*)*  (b).}
\end{table}

\begin{table}
\begin{tabular}{lcccc}
\hline
&\multicolumn{2}{c}{Difference} & \multicolumn{2}{c}{Relative Difference}\\
$L$(a.\ u.\ ) & $\alpha$ & $\gamma$  & $\alpha$ & $\gamma$\\
\hline
2.5    & -3.69  & 260.70  & -7.3 &  40.\\
3.0    & -1.47  & 33.8 & -5.4 & 37.\\
4.0    & -1.35  & 3.63  & -8.4 & 22. \\
\hline
\end{tabular}
\caption{\label{tab8}
Differences between  DMC and  HF values for the 
linear susceptibility  $\alpha$ in a.\ u.\ per H$_2$ unit and for the 
second hyper-susceptibility $\gamma$ in 10$^3$ a.\ u.\ for H$_2$ unit and 
relative differences in percent, for hydrogen chains
with bond-length alternations $L$. }
\end{table}

\begin{table}
\begin{tabular}{lcc}
\hline
$L$(a.\ u.\ ) & $\mathcal{E}$ (a.\ u.\ )& $\sigma^2$ (a.\ u.\ )\\
\hline
2.5  &  0.003 & 2.396 $\pm$ 0.009 \\
2.5  &  0.01  & 2.873 $\pm$ 0.035 \\
\hline
3.0  &  0.003 & 1.751 $\pm$ 0.005 \\
3.0  &  0.01  & 2.261 $\pm$ 0.040\\
\hline
4.0  &  0.01  & 1.391 $\pm$ 0.003\\
4.0  &  0.03  & 1.620 $\pm$ 0.014\\
\hline
\end{tabular}
\caption{\label{tab9}
DMC localization $\sigma^2$ along the longitudinal
direction of hydrogen chains with bond-length 
alternations $L$ for electric fields $\mathcal{E}$.}
\end{table}

\begin{table}
\begin{tabular}{lcc}
\hline
$N$ & $\alpha$ & $\gamma$ \\
\hline
10 &  16.04 $\pm$ 0.10 &  16.5 $\pm$ 1.1\\
20 &  15.70 $\pm$ 0.17 &  12.7 $\pm$ 2.9\\
\hline
\end{tabular}
\caption{\label{tab10}
DMC  linear susceptibility $\alpha$ in a.\ u.\ and second hyper-susceptibility $\gamma$ in units of
10$^3$ a.\ u.\ per H$_2$ unit for periodic linear hydrogen chains with
bond-length alternation $L$=4.0 a.\ u.\  calculated with supercells containing $N$  H$_2$ units.}

\end{table}

\begin{table*}
\begin{tabular}{lcccccc}
\hline
 $\mathcal{E}$ (a.\ u.\ )& $\Re\{\overline z_\mathrm{DMC}\}$ & $\Im\{\overline z_\mathrm{DMC}\}$ 
& $\mu_\mathrm{DMC}$ (a.\ u.\ )&
 $\Re\{\overline z_\mathrm{HF}\}$ & $\Im\{\overline z_\mathrm{HF}\}$ & $\mu_\mathrm{HF}$ (a.\ u.\ )\\
\hline
 0.    & 0.6146 $\pm$ 0.0011 & 0.1249 $\pm$ 0.0019 & 0.1436 $\pm$ 0.0019  & 0.6311 & 0.0000 & 0. \\
 0.003 & 0.6127 $\pm$ 0.0008 & 0.1341 $\pm$ 0.0013 & 0.1544 $\pm$ 0.0018  & 0.6118 & 0.1443 & 0.1654\\
 0.01  & 0.6102 $\pm$ 0.0011 & 0.1608 $\pm$ 0.0024 & 0.1846 $\pm$ 0.0029 & 0.4021 & 0.4536 & 0.6055\\
\hline
\end{tabular}
\caption{\label{tab11}
Longitudinal dipole moment $\mu$ and fixed point $\overline z$ calculated
through DMC and HF for an hydrogen chain with bond-length alternation
of 2.5 a.\ u.\ and for an electric field of 0.003 a.\ u.\
The trial wavefunctions are obtained from HF calculations
with electric fields $\mathcal{E}$.}
\end{table*}

\end{document}